\documentclass{PoS}

\title{SuperIso program and flavor data constraints}

\ShortTitle{SuperIso program and flavor data constraints}

\author{\speaker{Farvah Mahmoudi}%
\\
        High Energy Physics, Uppsala University, Box 535, 75121 Uppsala, Sweden\\
	LPC, Universit\'e Blaise Pascal, CNRS/IN2P3, 63177 Aubi\`ere Cedex, France\\
        E-mail: \email{mahmoudi@in2p3.fr}}

\abstract{
We present here an overview of the SuperIso program which is a public code dedicated to the calculation of flavor physics observables in supersymmetry. The main purpose of the SuperIso program is to provide, by confronting the theoretical predictions to the experimental data for flavor observables, indirect constraints on the parameters of supersymmetry. These constraints can then provide limits on physical masses of yet undiscovered new particles, such as charged Higgs bosons.
}

\FullConference{Prospects for Charged Higgs Discovery at Colliders\\
		 16-19 September 2008\\
		 Uppsala, Sweden}

\begin{document}
\section{Introduction}
Indirect searches appear as important tools in the quest for new physics and particles as they provide opportunities to study the flavor structure of new physics scenarios. Such information can therefore play a complementary role to the direct searches. The main purpose of the SuperIso program is to offer the possibility to evaluate the most important indirect observables in supersymmetry, which can be used to derive constraints on the model parameters and/or the new particle masses. In particular, SuperIso has been recently used to constrain the charged Higgs mass as reported in \cite{eriksson}.
\section{SuperIso program}
SuperIso \cite{superiso,superiso2} is a program dedicated mainly to the calculation of flavor physics observables in supersymmetry with minimal flavor violation. The calculation of the anomalous magnetic moment of the muon is also implemented in the program. SuperIso uses a SUSY Les Houches Accord file (SLHA1 or SLHA2) \cite{slha,slha2} as input, which can be either generated automatically by the program via a call to SOFTSUSY \cite{softsusy} or ISAJET \cite{isajet}, or provided by the user. The program is able to perform the calculations automatically for different supersymmetry breaking scenarios, such as the minimal Supergravity Grand Unification (mSUGRA), the Non-Universal Higgs Mass model (NUHM), the Anomaly Mediated Supersymmetry Breaking scenario (AMSB) and the Gauge Mediated Supersymmetry Breaking scenario (GMSB). 
\section{Flavor Observables}
A broad set of flavor physics observables is implemented in SuperIso. This includes isospin symmetry breaking in $B \to K^* \gamma$ decays, $b \to s \gamma$ branching ratio, the branching ratio of $B_s \to \mu^+ \mu^-$, the branching ratio of $B_u \to \tau \nu_\tau$, the branching ratios of $B \to D^0 \tau \nu_\tau$ and $B \to D^0 e \nu_e$, and the branching ratio of $K \to \mu \nu_\mu$. In the following we will briefly review each of these observables. The details of the calculations can be found in \cite{superiso2}.
\subsection{Inclusive branching ratio of $B \to X_s \gamma$}
The decay $B \to X_s \gamma$ proceeds through electromagnetic penguin loops, involving $W$ boson in the Standard Model, in addition to charged Higgs boson, chargino, neutralino and gluino loops in supersymmetric models. The contribution of neutralino and gluino loops is negligible in minimal flavor violating scenarios. The branching ratio of $B \to X_s \gamma$ can be written as \cite{misiak2}
\begin{equation}
\rm{BR}(\bar{B} \to X_s \gamma)= \mathrm{BR}(\bar{B} \to X_c e \bar{\nu})_{\rm exp} \left| \frac{ V^*_{ts} V_{tb}}{V_{cb}} \right|^2 \frac{6 \alpha}{\pi C} \left[ P(E_0) + N(E_0) + \epsilon_{em} \right] \;,
\end{equation}
with
\begin{equation}
C = \left| \frac{V_{ub}}{V_{cb}} \right|^2 \frac{\Gamma[\bar{B} \to X_c e \bar{\nu}]}{\Gamma[\bar{B} \to X_u e \bar{\nu}]} \;.
\end{equation}
$P(E_0)$ and $N(E_0)$ denote respectively the perturbative and non perturbative contributions, which involve the Wilson coefficients $C_{1\cdots 8}$, with $E_0$ a cut on the photon energy. $\epsilon_{em}$ is an electromagnetic correction. The calculation is performed at NNLO accuracy in SuperIso, with NNLO SM contributions and NLO SUSY corrections. 
\subsection{Isospin asymmetry of $B \to K^* \gamma$}
The isospin asymmetry $\Delta_{0}$ in $B \to K^* \gamma$ decays arises when the photon is emitted from the spectator quark. The contribution to the decay width depends therefore on the charge of the spectator quark and is different for charged and neutral $B$ meson decays:
\begin{equation}
\Delta_{0}=\displaystyle \frac{\Gamma(\bar B^0\to\bar K^{*0}\gamma) - \Gamma(B^\pm \to K^{*\pm}\gamma)}{\Gamma(\bar B^0\to\bar K^{*0}\gamma) + \Gamma(B^\pm\to K^{*\pm}\gamma)}\;,
\end{equation}
which can be written as \cite{kagan}:
\begin{equation}
\Delta_{0} =\mbox{Re}(b_d-b_u) \;,
\end{equation} 
where the spectator dependent coefficients $b_q$ take the form:
\begin{equation}
b_q = \frac{12\pi^2 f_B\,Q_q}{\overline{m}_b\,T_1^{B\to K^*} a_7^c}\left(\frac{f_{K^*}^\perp}{\overline{m}_b}\,K_1+ \frac{f_{K^*} m_{K^*}}{6\lambda_B m_B}\,K_{2q} \right)\;.
\end{equation}
The coefficients $a_7^c$, $K_1$ and $K_{2q}$ contain Wilson coefficients, at NLO accuracy in SuperIso. In the same way as for the $b \to s \gamma$ branching ratio, the SUSY contributions induced by charged Higgs and chargino loops must be taken into account for the calculation of isospin symmetry breaking.
\subsection{Branching ratio of $B_s\to \mu^+ \mu^-$}
The rare decay $B_s \to \mu^+ \mu^-$ proceeds via $Z$-mediated penguin and box diagrams in the SM, and the branching ratio is therefore highly suppressed. In supersymmetry, for large values of $\tan\beta$ this decay can receive large contributions from neutral Higgs bosons in chargino, charged Higgs and $W$-mediated penguins.
The branching fraction for $\mathrm{BR}(B_s \to \mu^+ \mu^-)$ is given by \cite{bobeth2,ellis}
\begin{eqnarray}
\mathrm{BR}(B_s \to \mu^+ \mu^-) &=& \frac{G_F^2 \alpha^2}{64 \pi^3} f_{B_s}^2 \tau_{B_s} M_{B_s}^3 |V_{tb}V_{ts}^*|^2 \sqrt{1-\frac{4 m_\mu^2}{M_{B_s}^2}} \nonumber \\
&\times& \left\{\left(1-\frac{4 m_\mu^2}{M_{B_s}^2}\right) M_{B_s}^2 | C_S |^2 + \left |C_P M_{B_s} -2 \, C_A \frac{m_\mu}{M_{B_s}} \right |^2\right\} \;.
\end{eqnarray} 
$C_A$ encloses the SM contributions \cite{ellis}:
\begin{equation}
C_A=\frac{1.033}{\sin^2 \theta_W}\left( \frac{\overline{m}_t(\overline{m}_t)}{170 \, \rm{GeV}} \right)^{1.55}\;,
\end{equation} 
which includes both LO and NLO QCD corrections. $C_S$ and $C_P$ encompass the SUSY loop effects due to Higgs boson contributions, box and penguin SUSY diagrams and counter-terms. They can receive large contributions, in particular in the high $\tan\beta$ regime. 
\subsection{Branching ratio of $B_u \to \tau \nu$}
The purely leptonic decay $B_u \to \tau \nu_\tau$ occurs via $W^+$ and $H^+$ mediated annihilation processes. This decay is helicity suppressed in the SM, but there is no such suppression for the charged Higgs exchange at high $\tan\beta$, and the two contributions can therefore be of similar magnitudes. This decay is thus very sensitive to charged Higgs boson and provide important constraints.

The branching ratio of $B_u \to \tau \nu_\tau$ reads \cite{hou-akeroyd}
\begin{equation}
\mathrm{BR}(B_u\to\tau\nu_\tau)=\frac{G_F^2f_B^2|V_{ub}|^2}{8\pi}\tau_B m_B m_\tau^2\left(1-\frac{m_\tau^2}{m_B^2}\right)^2 \left[1-\left(\frac{m_B^2}{m_{H^+}^2}\right)\frac{\tan^2\beta}{1+\epsilon_0\tan\beta}\right]^2 \;,
\end{equation}
where $\epsilon_0$ corresponds to a two loop SUSY correction. The ratio 
\begin{equation}
R^{\mathrm{MSSM}}_{\tau\nu_\tau}=\frac{\mathrm{BR}(B_u\to\tau\nu_\tau)_{\mathrm{MSSM}} }{\mathrm{BR}(B_u\to\tau\nu_\tau)_{\mathrm{SM}}}=\left[1-\left(\frac{m_B^2}{m_{H^+}^2}\right)\frac{\tan^2\beta}{1+\epsilon_0\tan\beta}\right]^2 \;,
\end{equation}
is usually considered to express the new physics contributions.
\subsection{Branching ratio of $B\to D \tau \nu$}
The semileptonic decay $B \to D \tau \nu_\tau$ is similar to $B_u \to \tau \nu_\tau$. The SM helicity suppression here occurs only near the kinematic endpoint. The branching ratio of $B \to D \tau \nu_\tau$ on the other hand is about 50 times larger that the branching ratio of $B_u \to \tau \nu_\tau$ in the SM.
The partial rate of the transition $B\to D\ell \nu_\ell$ can be written as \cite{hou,kamenik}
\begin{eqnarray}
\frac{d\Gamma(B\to D\ell \nu_\ell)}{dw} &=& \frac{G_F^2|V_{cb}|^2 m_B^5}{192\pi^3}\rho_V(w) \label{partialbdtaunu}\\
&& \times\left[1 - \frac{m_{\ell}^2}{m_B^2}\, \left\vert 1- t(w)\, \frac{m_b}{(m_b-m_c)m^2_{H^{+}}}\,\frac{\tan^2\beta}{1+\epsilon_0\tan\beta} \right\vert^2 \rho_S(w) \right]\;,\nonumber 
\end{eqnarray}
where $w=v_D\cdot v_B$ is a kinematic variable and $t(w) = m_B^2+ m_D^2 - 2 w m_D m_B$. $\rho_V$ and $\rho_S$ are the vector and scalar Dalitz density contributions \cite{kamenik}.
The branching fraction can then be obtained by integrating Eq. (\ref{partialbdtaunu}).
The following ratio
\begin{equation}
\xi_{D\ell\nu} = \frac{\mathrm{BR}(B\to D^0\tau\nu_\tau)}{\mathrm{BR}(B\to D^0 e \nu_e)}
\end{equation}
is also considered in order to reduce some of the theoretical uncertainties.
\subsection{Branching ratio of $K\to \mu \nu$}
The leptonic kaon decay $K \to \mu \nu$ is also mediated via $W^+$ and $H^+$ annihilation processes. The charged Higgs contribution in this case is reduced since $H^+$ couples to lighter quarks. In SuperIso we consider the following ratio in order to reduce the theoretical uncertainties from $f_K$ \cite{flavianet}
\begin{eqnarray}
\displaystyle\frac{\rm{BR}(K \to \mu \nu_\mu)}{\rm{BR}(\pi \to \mu \nu_\mu)}&=& 
\frac{\tau_K}{\tau_\pi}\left|\frac{V_{us}}{V_{ud}} \right|^2 \frac{f^2_K}{f^2_\pi} \frac{m_K}{m_\pi}\left(\frac{1-m^2_\ell/m_K^2}{1-m^2_\ell/m_\pi^2}\right)^2 \nonumber \\
&& \times \left[1-\frac{m^2_{K^+}}{M^2_{H^+}}\left(1 - \frac{m_d}{m_s}\right)\frac{\tan^2\beta}{1+\epsilon_0\tan\beta}\right]^2 \left(1+\delta_{\rm em}\right)\;,
\end{eqnarray}
where $\delta_{\rm em} = 0.0070 \pm 0.0035$ is a long distance electromagnetic correction factor. The quantity $R_{\ell 23}$ \cite{flavianet} is also implemented in SuperIso:
\begin{equation}
R_{\ell 23}=\left| \frac{V_{us}(K_{\ell 2})}{V_{us}(K_{\ell 3})} \times \frac{V_{us}(0^+ \to 0^+)}{V_{ud}(\pi_{\ell 2})} \right|=\left|1-\frac{m^2_{K^+}}{M^2_{H^+}}\left(1 - \frac{m_d}{m_s}\right)\frac{\tan^2\beta}{1+\epsilon_0\tan\beta}\right|\;.
\end{equation}
\section{Anomalous magnetic moment of muon}
The discrepancy between the experimental value and the SM prediction for the anomalous magnetic moment of the muon $a_\mu = (g-2)/2$ 
can be explained by supersymmetric contributions through chargino-sneutrino and neutralino-smuon loops \cite{martin}. In SuperIso, the one-loop SUSY contributions to $a_\mu$ are implemented, as well as the suppression due to the leading logarithm contribution from two-loop evaluation.
\section{Constraints}
\begin{figure} 
\begin{center}
\includegraphics[width=8cm]{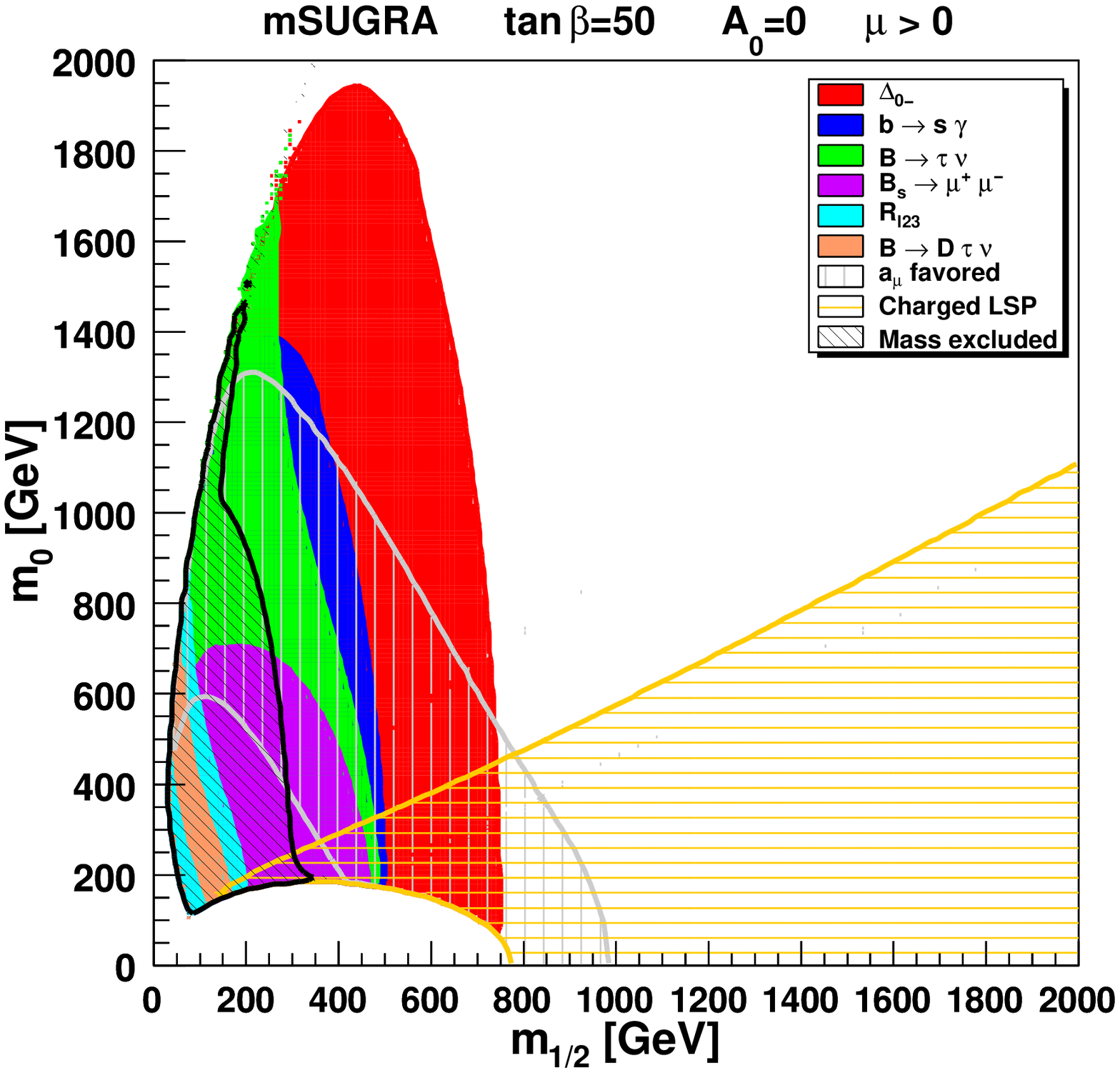}\hspace*{-0.1cm} \includegraphics[width=8cm]{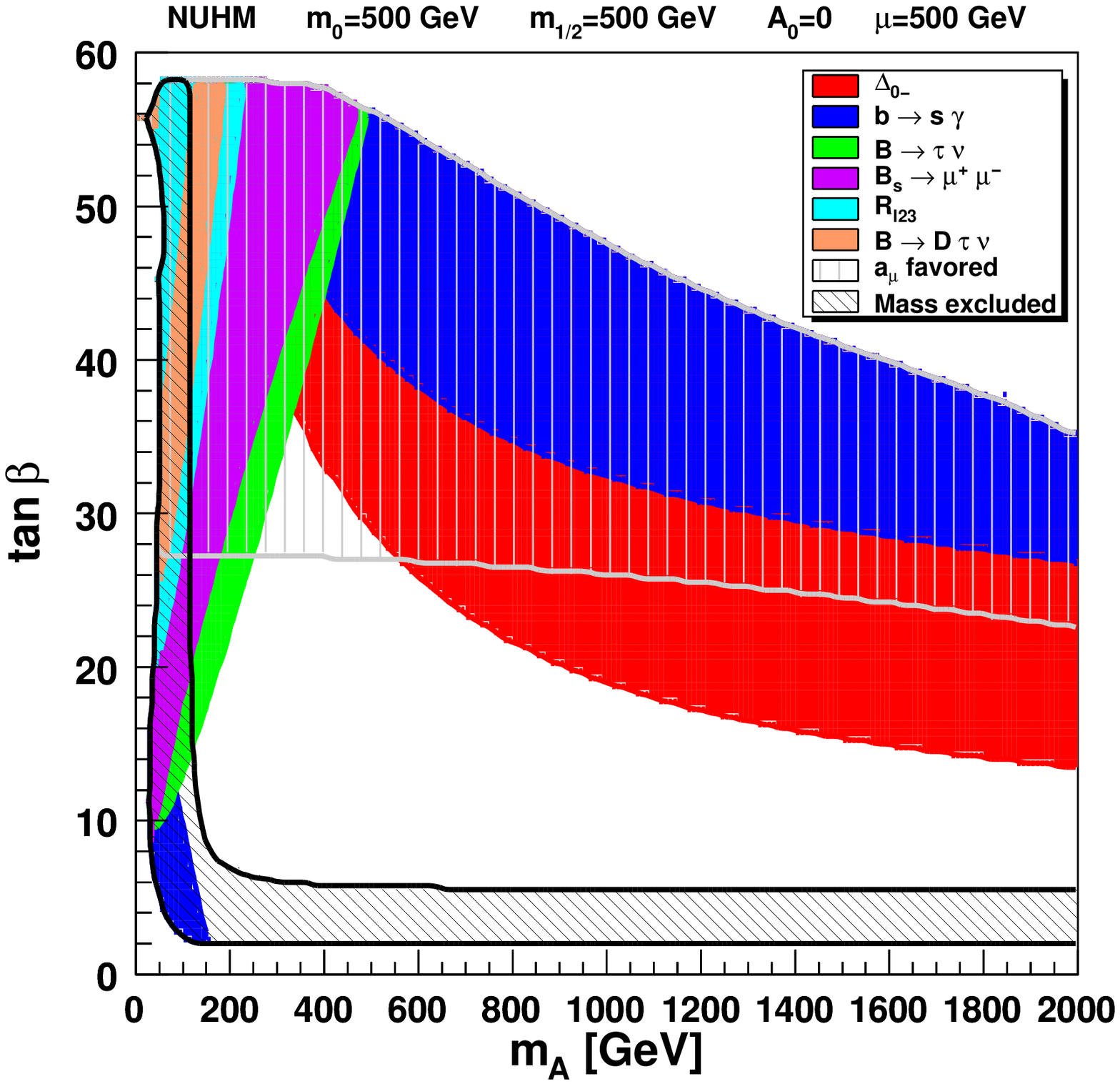}
\caption{Constraints in mSUGRA $(m_{1/2} - m_0)$ parameter plane (left) and in NUHM $(m_A - \tan\beta)$ parameter plane (right). The contours are superimposed in the order given in the legend.\vspace*{-0.4cm}}
\label{fig1}
\end{center}
\end{figure}

Constraints can be obtained using SuperIso by comparing, for each SUSY parameter point, the calculated values for the above observables with the experimental allowed intervals \cite{eriksson,mahmoudi}. We show two examples of results in figure~\ref{fig1} in the mSUGRA and NUHM parameter spaces. The colored regions in this figure correspond to the regions excluded at 95\% C.L. by the observables except for $a_\mu$ which corresponds to the favored region. 
Since charged Higgs bosons appear in most of the aforementioned observables, and in particular at tree level already in leptonic and semileptonic decays, it is possible to impose severe constraints on the MSSM Higgs sector, as shown in \cite{eriksson}. One should however not over-interpret these results as the observables are subject to uncertainties mainly from flavor parameters. They provide nevertheless useful information as can be seen from figure~\ref{fig1}.
\section{Conclusion}
SuperIso is able to compute numerous flavor physics observables -- as well as the muon anomalous magnetic moment -- which can be used to explore and constrain the SUSY parameter space. They can be used also to constrain the physical masses, and in particular of the charged Higgs boson since it appears as an additional mediator of charged current interactions.

New features will continue to appear in the next versions of SuperIso, and in particular new flavor observables and new supersymmetric scenarios will be added.\vspace*{-0.5cm}
\section*{}

\end{document}